\documentclass[final]{cvpr}

\usepackage[accsupp]{axessibility}  

\usepackage{float}
\usepackage{times}
\usepackage{epsfig}
\usepackage{graphicx}
\usepackage{amsmath}
\usepackage{amssymb}

\usepackage[pagebackref,breaklinks,colorlinks]{hyperref}

\def\cvprPaperID{3138} 
\def\confName{CVPR}
\def\confYear{2023}
\graphicspath{{./figs/}{./figs/result/}}

\begin{document}

\title{Image Quality-aware Diagnosis via Meta-knowledge Co-embedding}

\author{
 Haoxuan Che \hspace{1.0cm} Siyu Chen\hspace{1.0cm} Hao Chen\footnotemark[1] \hspace{1.0cm}\\
  The Hong Kong University of Science and Technology \hspace{0.5cm} \\
\texttt{\footnotesize \{hche,schende,jhc\}@cse.ust.hk}
}
 

\maketitle
\pagestyle{empty}  
\thispagestyle{empty} 

\renewcommand{\thefootnote}{\fnsymbol{footnote}} 
\footnotetext[1]{\noindent Corresponding author: Hao Chen, email: jhc@cse.ust.hk.}

\begin{abstract}
Medical images usually suffer from image degradation in clinical practice, leading to decreased performance of deep learning-based models.
To resolve this problem, most previous works have focused on filtering out degradation-causing low-quality images while ignoring their potential value for models.
Through effectively learning and leveraging the knowledge of degradations, models can better resist their adverse effects and avoid misdiagnosis.
In this paper, we raise the problem of \textbf{image quality-aware diagnosis}, which aims to take advantage of low-quality images and image quality labels to achieve a more accurate and robust diagnosis.
However, the diversity of degradations and superficially unrelated targets between image quality assessment and disease diagnosis makes it still quite challenging to effectively leverage quality labels to assist diagnosis.
Thus, to tackle these issues, we propose a novel \textbf{meta-knowledge co-embedding network}, consisting of two subnets: Task Net and Meta Learner.
Task Net constructs an explicit quality information utilization mechanism to enhance diagnosis via knowledge co-embedding features, 
while Meta Learner ensures the effectiveness and constrains the semantics of these features via meta-learning and joint-encoding masking.
Superior performance on five datasets with four widely-used medical imaging modalities demonstrates the effectiveness and generalizability of our method.
\end{abstract}

\vspace{-2mm}
\section{Introduction}
\label{sec:intro}
Medical imaging is one of the most valuable sources of diagnostic information about anatomical structures and pathological characteristics \cite{liu2019comparison}.
Advanced deep learning-based methods applied to high-quality (HQ) medical images have shown significant potential in disease analysis and diagnosis \cite{li2021applications,jiang2022deep}, achieving favorable results compared with human healthcare professionals \cite{nagendran2020artificial}.
However, in clinical practice, obtaining HQ images is not always feasible.
Medical images often exhibit significant variations in imaging quality due to factors such as patient movements or environmental conditions \cite{liu2022deepdrid,holmen2020prevalence}.
For instance, a medical image quality assessment study of 28,780 fundus images revealed that approximately 41.6\% of them contained image artifacts and corruption and were considered low-quality (LQ) \cite{fu2019evaluation}.
Such degradations can increase the uncertainty in pathological observation, leading to misdiagnosis \cite{beede2020human, liu2022understanding}.

\begin{figure}[t]
    \centering
    \includegraphics[width=\linewidth]{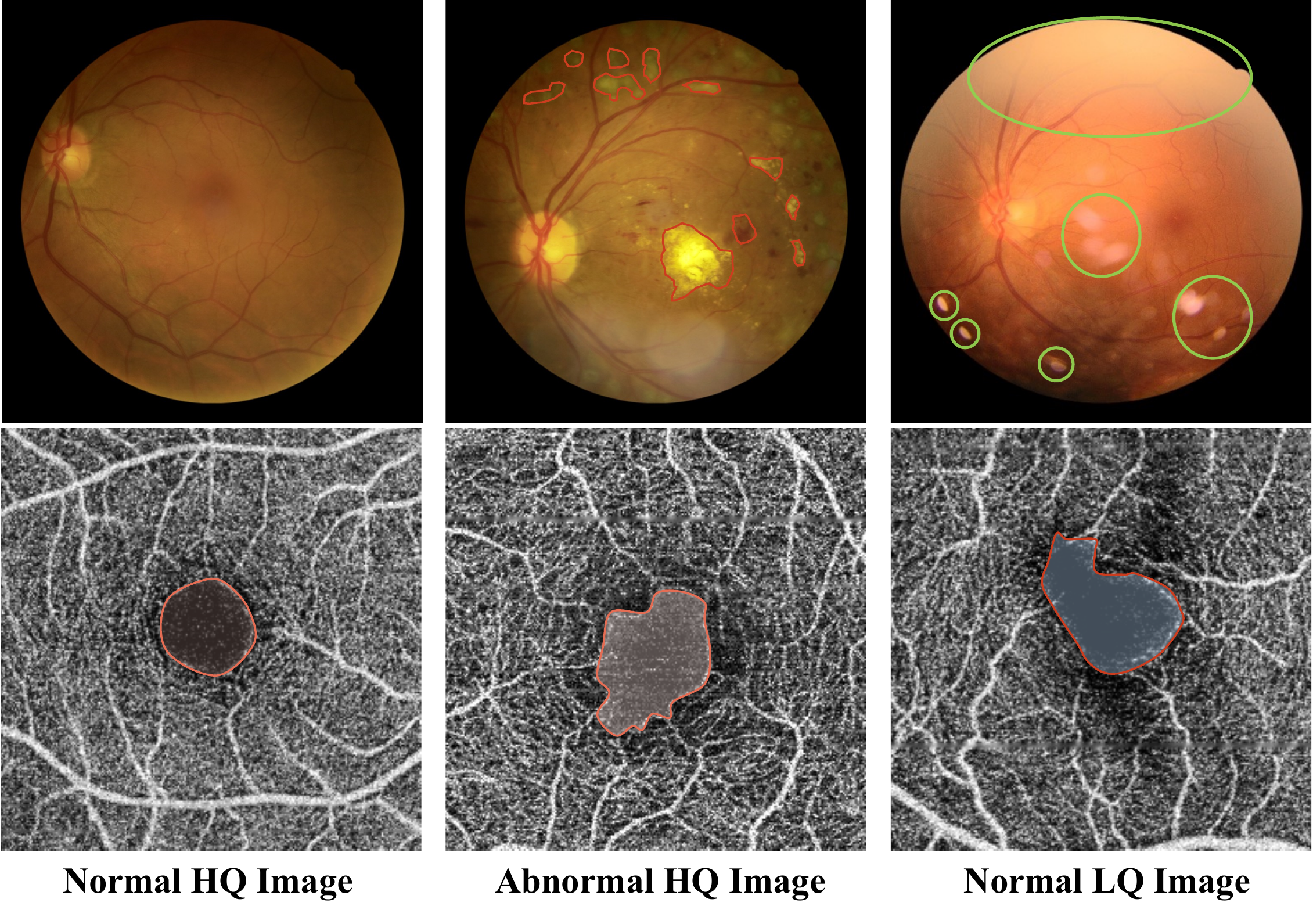}
    \caption{Illustration of impact of image degradations on diagnosis semantics for fundus (top) and OCTA (bottom) images. 
    \textbf{Top}: Degradations obscure parts of the vessel structure and present lesion-like spots.
    \textbf{Bottom}: Degradations result in a fake enlargement of the foveal avascular zone (central circular area). 
    }
    \label{fig:1}
 \vspace{-4mm}
\end{figure}

Medical image degradations can significantly affect diagnostic semantics, as illustrated in Figure \ref{fig:1}.
For instance, shadow degradation can obscure anatomical structures crucial for diagnosis, while spot artifacts can obfuscate pathological signs that typically manifest as circular shapes \cite{shen2020modeling}. 
Furthermore, image degradations can also affect diagnostic measurements, such as the vessel area density in optical coherence tomography angiography (OCTA) images, rendering them unreliable \cite{holmen2020prevalence}. 
These close relationships raise challenges in distinguishing degradations from actual abnormalities\cite{shen2020modeling}, leading to false knowledge of lesions and undesired misdiagnosis during training and deployment \cite{liu2022understanding}. 
Aware of the profound influence of image quality on diagnosis, many previous works have focused on utilizing image quality assessment to select relatively HQ images for training and testing, thereby avoiding the influence of LQ images \cite{fu2019evaluation,ziqi2022using,shen2020domain,hu2022chest,zhou2020retinal}.
However, discarding LQ images containing diagnostically valuable information results in a waste of precious clinical data \cite{fu2019evaluation}. 
Including LQ images and corresponding quality information in training can assist models in recognizing potential false abnormalities, thus achieving more robust and accurate diagnosis \cite{yii2022rethinking,zhou2018fundus,luo2020deep}.

In this paper, we reconsider the value of LQ images and corresponding image quality labels, and introduce the problem of \textit{image quality-aware diagnosis} (IQAD). 
The goal of IQAD is to enable models to leverage LQ images while simultaneously learning image quality labels to achieve an accurate and robust diagnosis.
However, effectively leveraging quality labels for diagnosis is non-trivial with a multi-task learning framework. 
Specifically, image quality assessment can be considered as a task ``unrelated'' to disease diagnosis \cite{ruder2017overview}, since it focuses on capturing image degradations, while diagnosis emphasizes identifying lesions. 
This distinction makes it challenging for models to effectively utilize image quality labels. 
Further, commonly-used coarse annotations of quality may not sufficiently reflect the diversity of image degradation, making it difficult to provide information that could be useful for precise diagnosis.

To achieve IQAD, we propose a novel \textit{meta-knowledge co-embedding network} (MKCNet) consisting of two subnets, Task Net and Meta Learner.
To enable leveraging potential benefits of quality information, Task Net conducts diagnosis predictions by explicitly leveraging knowledge co-embedding features with desired knowledge of image quality and disease diagnosis.
These features are constructed by learning auxiliary label embeddings from Meta Learner. 
Further, we employ meta-learning and joint-encoding masking to ensure the effectiveness and semantics of auxiliary label embedding and circumvent the barrier of obtaining fine-grained image quality labels.
Specifically, joint-encoding masking selects a part of the Meta Learner output as auxiliary label embedding through combinations of quality and diagnosis labels.
Moreover, Meta Learner learns to provide auxiliary label embedding in a meta-learning manner to assist Task Net optimization, encouraging it to learn effective knowledge co-embedding features. 
Our main contributions are highlighted as follows:

(1) We tackle a novel problem named IQAD. To the best of our knowledge, this is the first work to discuss and analyze this critical and practical problem.

(2) We propose a novel method, MKCNet, to effectively handle the challenges posed by annotation granularity and task focus discrepancy via leveraging quality information explicitly and introducing a meta-learning paradigm. 

(3) We conduct extensive experiments on five datasets spanning four different yet widely-used medical imaging modalities. Our in-depth analytical study demonstrates the effectiveness and generalizability of MKCNet.


\begin{figure*}[ht]
  \centering
    \includegraphics[width=\linewidth]{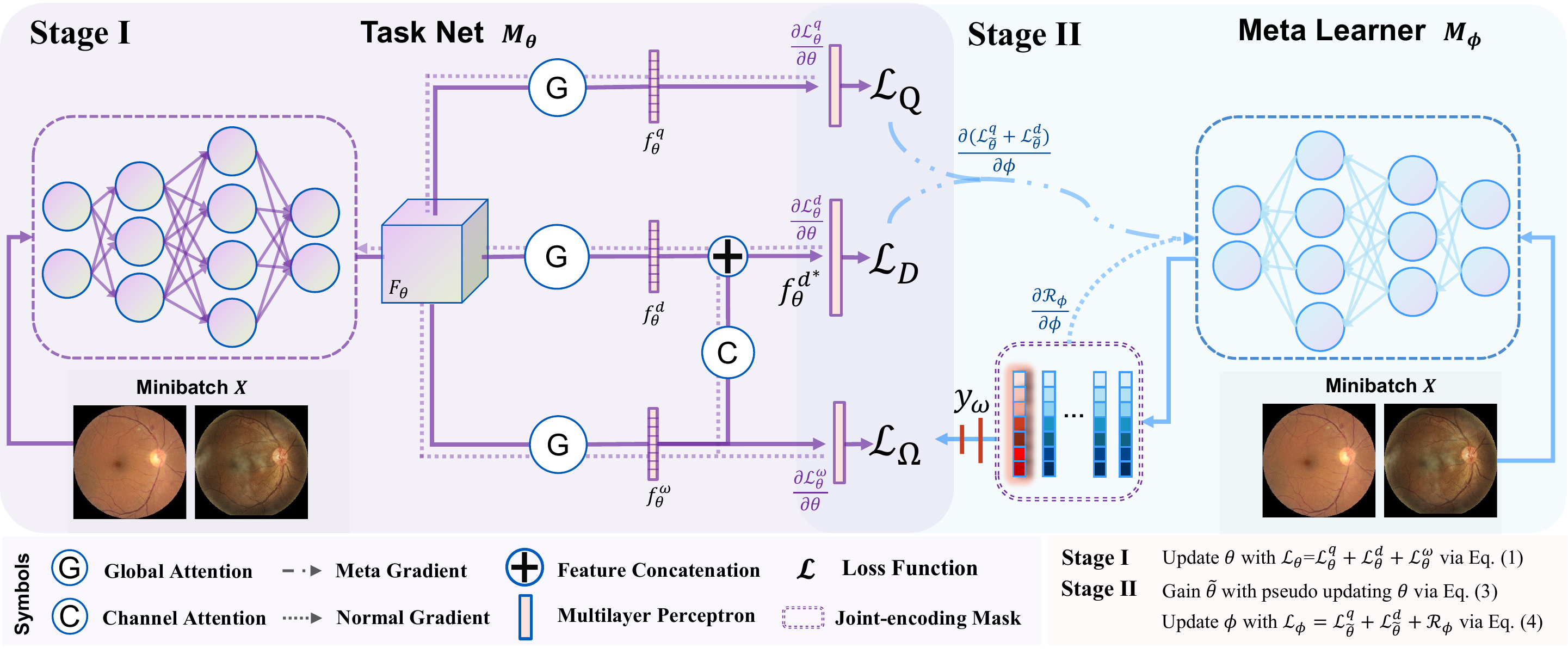}
    \caption{The overview of MKCNet with two subnets (Task Net $\mathcal{M}_\theta$ and Meta Learner $\mathcal{M}_\phi$).
    In the first stage, $\mathcal{M}_\theta$ learns to construct $f_\theta^\omega$ by leveraging $y_\omega$. 
    Simultaneously, it adopts global attention to learn an informative and generalizable $F_\theta$, while it explicitly utilizes $f_\theta^\omega$ and $f_\theta^d$ to make diagnoses.
    In the second stage, $\mathcal{M}_\phi$ learns to provide $y_\omega$ with desired knowledge of image quality and disease diagnosis. 
    $\mathcal{M}_\phi$ ensures the effectiveness of $f_\theta^\omega$ while constraining its semantics by utilizing the joint-encoding masking and meta-optimization.
    }
    \label{fig:pipeline}
\vspace{-3mm}
\end{figure*}

\section{Related Work}
\textbf{Disease diagnosis.}
Many deep-learning methods have been developed to diagnose diseases \cite{he2020cabnet,liu2020green,wang2023deep,wang2019weakly,wang2021deep,bai2022transformer,luo2022pseudo}.
For example, He et al. \cite{he2020cabnet} proposed CABNet for learning discriminative features associated with different severities of diabetic retinopathy (DR), and Liu et al. \cite{liu2020green} developed a convolutional graph networks-based method to explore potential relationships among grades of DR.
However, these methods may not perform well when dealing with image degradations as they do not consider image quality issues \cite{liu2022understanding}.
By contrast, MKCNet leverages LQ images and quality labels to achieve a more robust and accurate diagnosis.

\textbf{Quality assessment and image enhancement.}
Aiming at avoiding the effect of LQ images, several methods have been proposed to assess image quality to select relatively HQ images \cite{fu2019evaluation,zhou2020retinal,ziqi2022using}.
For example, Fu et al. \cite{fu2019evaluation} selected usable samples and rejected nearly 20\% of images as unsuitable for model learning.
However, these images would still be diagnosable for physicians.
Rejecting diagnostically valuable LQ images is wasteful, since LQ images are not only useful in model generalization ability improvement in training \cite{geirhos2018imagenet,yii2022rethinking}, but also effective in evaluating the robustness of models in testing \cite{ovadia2019can,geirhos2020shortcut}.
Another solution is to leverage generative models to enhance the quality of LQ images \cite{liu2022degradation,yang2023retinal,deng2022rformer, zhang2022frequency}.
For example, Shen et al. \cite{shen2020modeling} improve the quality of LQ fundus images by generating degradations on images to simulate pseudo-paired samples.
However, it is costly to train such generative models, which requires a large number of images for desirable performance \cite{zhang2022frequency}.
Additionally, the simulation may only represent a partial distribution of realistic degradation \cite{romero2022unpaired}, and it is rarely helpful in improving recognition performance \cite{liu2022exploring}.
Meanwhile, capturing real pairs with different image qualities proves to be extremely challenging \cite{yang2019bi}.
Instead of requiring a high cost, MKCNet leverages the multi-task learning framework to tackle the IQAD problem in a less costly and more effective manner.

\textbf{Multi-task learning.}
It has been shown that multi-task learning is an effective method for training a generalized model that can simultaneously handle multiple tasks \cite{ruder2017overview}. 
In the medical domain, many studies have used it to improve the performance of models by exploring internal relationships among diseases \cite{li2019canet,che2022learning,liu2019multi,chen2019multi,bai2021influence}, as well as utilizing auxiliary tasks to assist primary tasks \cite{lin2021seg4reg+,hsieh2021boosting}. 
However, most of these works ignore the potential benefits of quality information on diagnosis, which is critical for the IQAD problem.
Among them, Zhou et al. \cite{zhou2018fundus} treated quality assessment as an auxiliary task.
However, they did not obtain a significant performance boost due to the absence of an explicit mechanism for utilizing quality information.
By contrast, MKCNet explicitly models and explores the potential assistance of quality information in diagnosis.

\section{Methodology}

This section first provides an introduction to IQAD and presents a preliminary experiment, as well as highlights the challenges involved. 
We then introduce our proposed solution, MKCNet, which explicitly addresses these challenges. 
Figure \ref{fig:pipeline} shows an overview of MKCNet.

\subsection{Image Quality-aware Diagnosis}
\label{sec:analysis}
\textbf{Preliminaries.} 
Given an image $x \in \mathcal{X}$, $y_d \in \mathcal{Y}_D$ and $y_q \in \mathcal{Y}_Q$ denote the corresponding disease diagnosis and image quality labels, respectively.
The target of IQAD is to train a model $\mathcal{F}: \mathcal{X} \rightarrow \mathcal{Y}_D$ to achieve robust and accurate diagnosis by leveraging both HQ and LQ images together with corresponding diagnosis and quality labels.

\begin{figure}[t]
    \centering
    \includegraphics[width=\linewidth]{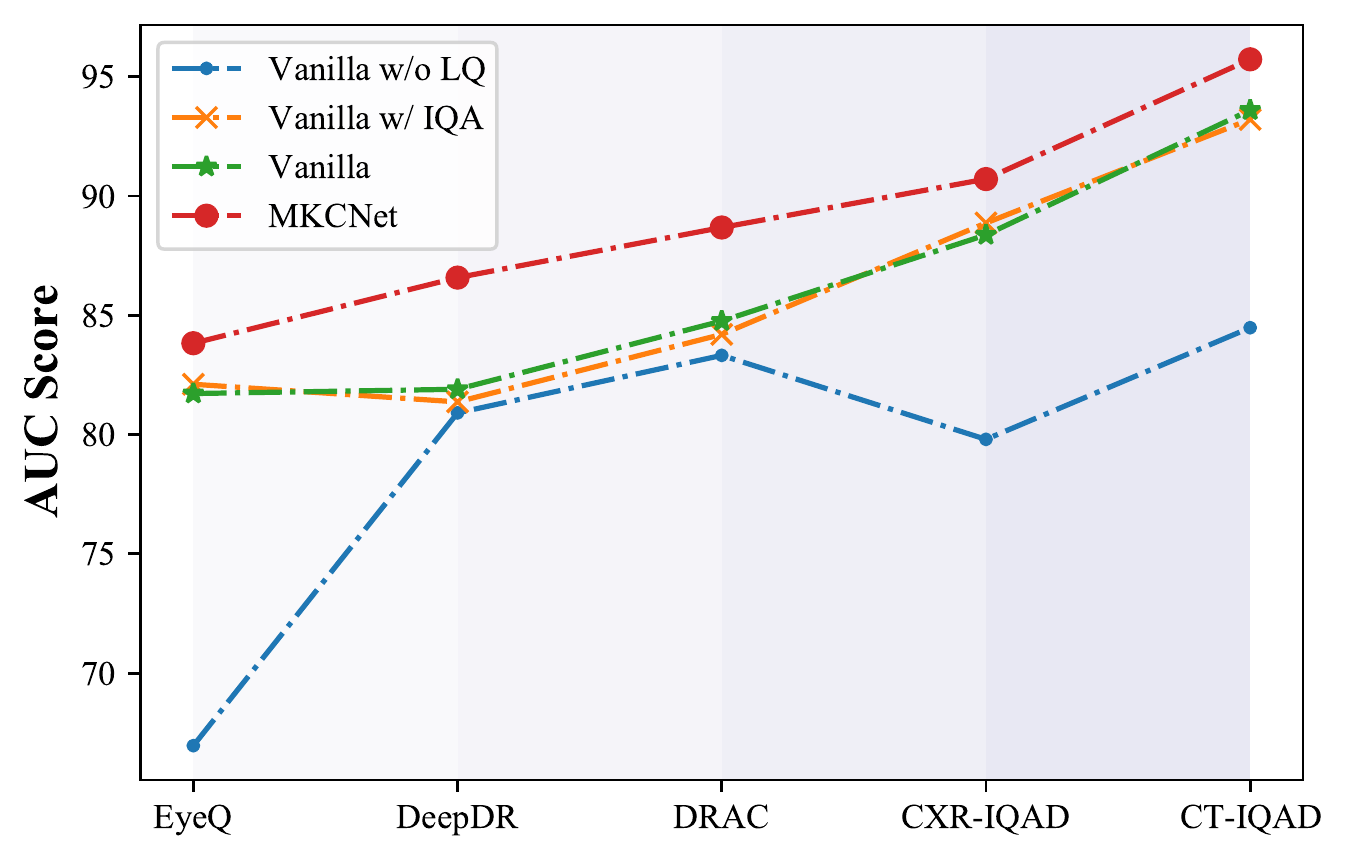}
    \caption{Influence of LQ images and an IQA module.
        It is useful for Vanilla to leverage LQ images, but
        the IQA module improves only marginally or harms performance.
        MKCNet can leverage LQ images and quality labels more effectively than Vanilla.}
    \label{fig:auc}
\vspace{-3mm}
\end{figure}

Intuitively, treating image quality assessment (IQA) as an auxiliary branch in a multi-task learning framework may seem like a straightforward solution. 
Thus, to explore this approach, we conduct a preliminary experiment on a VGG16 model (denoted as Vanilla) to assess the impact of learning LQ images and leveraging quality labels, as shown in Figure \ref{fig:auc}.
As expected, leveraging LQ images is beneficial for model learning.
Unexpectedly, however, we observed that incorporating quality labels only marginally enhances or even hinders the diagnostic performance, despite the fact that these labels may imply false abnormalities and provide extra useful diagnostic information. 
It suggests that effectively leveraging quality labels is non-trivial.

\textbf{Challenges.} 
In the context of multi-task learning, we argue that the challenges of IQAD are two-fold. 
\textbf{The first challenge} arises from the non-direct relationship between image quality and diagnosis, which makes it difficult for models to associate them without an explicit mechanism. 
Generally, models trained for disease diagnosis focus on lesion areas and anatomical structures, while image quality assessment requires models to capture image degradations.
This discrepancy in the task focus necessitates a specialized design that enables models to effectively leverage quality information.
To address this challenge, we design Task Net, which includes an explicit utilization mechanism to leverage knowledge co-embedding features for diagnosis.
\textbf{The second challenge} arises due to the limited granularity of binary or multi-class quality annotations, which fail to capture the diverse types and degrees of image degradations \cite{shen2020modeling,liu2022understanding}. 
It results in a lack of detailed information that can guide the model and makes it challenging to establish fixed patterns of how image quality affects disease diagnosis \cite{luo2022rethinking}. 
Additionally, determining image-level label criteria to associate quality and diagnosis is complex, and annotating images with pixel-level information of degradations is costly \cite{hu2022chest}. 
Due to label limitations, even if an explicit utilization mechanism exists, the model may not effectively leverage quality information.
To tackle this issue, we adopt a meta-learning paradigm, where the Meta Learner is trained to provide auxiliary label embeddings that present adaptive correlations between image quality and disease diagnosis labels, allowing the model to effectively utilize quality information.

\subsection{Meta-knowledge Co-embedding Network}
As shown in Figure \ref{fig:pipeline}, MKCNet is composed of two subnets, Task Net denoted as $\mathcal{M}_{\theta}$ with parameter $\theta$, and Meta Learner denoted as $\mathcal{M}_{\phi}$ with parameter $\phi$. 
Given an image $x$ and its corresponding image quality and disease diagnosis labels $y_q$ and $y_d$, we obtain an auxiliary label embedding $y_\omega$ from the output vector $\mathcal{M}_{\phi}(x)$ of Meta Learner. 
The corresponding predictions of Task Net for $y_q$, $y_d$, and $y_\omega$ are denoted as $\mathcal{M}_{\theta}^q(x)$, $\mathcal{M}_{\theta}^d(x)$, and $\mathcal{M}_{\theta}^\omega(x)$, respectively.

Inspired by the recent success of meta-learning \cite{li2018learning,liu2019self}, we adopt a two-stage learning paradigm to optimize MKCNet.
In the first stage, Task Net is trained with $y_q$, $y_d$, and $y_\omega$, and it explicitly leverages knowledge co-embedding features to improve diagnosis performance.
In the second stage, Meta Learner learns to provide $y_\omega$ via meta-learning, to optimize knowledge co-embedding features in Task Net.
The two stages are iterative during each training epoch, leading to end-to-end interaction between the two subnets.

\vspace{-2mm}
\subsubsection{Task Net for Quality-aware Diagnosing}
To address the first challenge discussed earlier, Task Net is designed to construct knowledge co-embedding features while explicitly exploring their potential usefulness in improving diagnosis.
To achieve this, we design the global attention block (GAB) in Task Net to extract task-specific features and the meta-knowledge assistance block (MAB) to explicitly explore their potential benefits in diagnosis.
In Task Net, we denote the feature map from the backbone as $F_\theta$, the features from three GABs as $f_{\theta}^q$, $f_{\theta}^d$, and $f_{\theta}^\omega$, and the feature from the MAB for final diagnosis as $f_{\theta}^{d^*}$.

\textbf{Knowledge co-embedding feature.} 
Task Net constructs the knowledge co-embedding feature $f_{\theta}^\omega$ by learning $y_\omega$, which contains the desired information beneficial for both image quality and disease diagnosis. 
To achieve this, it first needs to learn informative and generalizable $F_\theta$, encompassing comprehensive semantics,
and then to construct the $f_{\theta}^\omega$ with applicable information for quality-aware diagnosis from $F_\theta$.
Therefore, We designed three learning branches with GAB in Task Net, which receive supervision signals $y_q$, $y_d$, and $y_\omega$, respectively. 
Intuitively, learning $y_q$ and $y_d$ helps Task Net to learn a joint feature space of image quality and disease diagnosis, leading to more informative $F_\theta$ with multiple semantics \cite{paredes2012exploiting}. 
Additionally, we adopt GAB to conduct both channel- and spatial-wise attention to highlight the potentially correlated features from $F_\theta$ to each supervision signal\cite{woo2018cbam}.
GABs extract features specific to $y_q$ and $y_d$ to encourage the construction of generalizable $F_\theta$ \cite{li2019canet}, while filtering irrelevant channels and spatiality to capture desired $f_{\theta}^\omega$ correlated with $y_\omega$.
Finally, Task Net obtains $f_{\theta}^d$ and $f_{\theta}^\omega$ containing diagnosis-related information and desired knowledge of image quality and disease diagnosis, respectively, for further explicit utilization.

\textbf{Meta-knowledge assistance block.} 
In addition, we designed the MAB to explicitly explore the potential assistance of $f_{\theta}^\omega$ for disease diagnosis, where $f_{\theta}^\omega$ learns joint semantics of disease diagnosis and image quality. 
MAB first filters channels of $f_{\theta}^\omega$ uncorrelated with disease diagnosis via a channel-wise attention block to ensure the desired functionality of $f_{\theta}^\omega$. 
It then utilizes the concatenated feature $f_{\theta}^{d^*}$ of $f_{\theta}^{d}$ and a filtered $f_{\theta}^\omega$ for the diagnosis. 
This ``filter and utilize'' pattern is critical for quality-aware diagnosis since $f_{\theta}^\omega$ may contain undesired or even misleading information.
Furthermore, the explicit utilization encourages the models to be aware of the potential benefits and assistance of image quality in the diagnosis.
Finally, the objective function $\mathcal{L}_\theta$ of Task Net $\mathcal{M}_{\theta}$ is denoted as
\vspace{-1mm}
\begin{equation}
\label{eq1}
\begin{split}
    \mathcal{L}_\theta = \mathcal{L}_D (\mathcal{M}_{\theta}^d (x), y_d)+ \mathcal{L}_Q (\mathcal{M}_{\theta}^q (x), y_q) \\ 
    + \mathcal{L}_\Omega (\mathcal{M}_{\theta}^\omega (x), y_\omega) \text{,}
\end{split}
\end{equation}
where $\mathcal{L}_D$, $\mathcal{L}_Q$ and $\mathcal{L}_\Omega$ represent the loss functions for disease diagnosis, image quality prediction, and knowledge co-embedding learning, respectively.
We would like to clarify that our focus is on the design philosophy rather than specific implementation details, and our proposed design is compatible with any attention modules such as \cite{qin2021fcanet,roy2018recalibrating}. 
We argue that a practical and appropriate design is the key to solving the IQAD problem. 
In this paper, we adopt a classic attention module CBAM \cite{woo2018cbam} rather than other modern attention modules to demonstrate the effectiveness of our design. 
Additionally, we conduct experiments to explore the significance of the design of Task Net, i.e., the existence of GAB and MAB, in the experiment section.

\vspace{-2mm}
\subsubsection{Meta Learner to Co-embed Knowledge}
As discussed in Section \ref{sec:analysis}, the current quality labels do not capture the diversity of degradations and their corresponding impact on diagnostic semantics.
We overcome this limitation by Meta Learner, which constrains the semantics and ensures the effectiveness of $f_{\theta}^\omega$ via \textit{joint-encoding masking} and \textit{meta-auxiliary learning}.

\textbf{Joint-encoding masking.}
We design the joint-encoding masking approach to ensure the effectiveness and semantic constraints of $f_{\theta}^\omega$ by constraining Meta Learner to provide a discriminative $y_\omega$ for both quality and diagnosis labels. 
To achieve this, we select $y_\omega$ from $\mathcal{M}_{\phi}(x)$ based on a joint encoding $y_{d,q}$ of $y_d$ and $y_q$, and create a binary mask $\mathcal{B}{y_{d,q}}$ that assigns a value of 1 to positions related to $y_{d,q}$ and 0 to others. 
Consider the case that both $y_q \text{ and } y_d \in \{ 0, 1 \}$, and the length of $\mathcal{M}_{\phi}(x)$ is 4.
In this case, the joint label encoding will be $y_{d,q} \in \{ 00, 01, 10, 11 \}$, and $\mathcal{B}_{y_{d,q}}$ will be $[1,0,0,0]$, $[0,1,0,0]$, $[0,0,1,0]$ and $[0,0,0,1]$, respectively.
Further, $y_\omega$ is derived from $\mathcal{M}_{\phi}(x)$ as follows:

\vspace{-1mm}
\begin{equation}
    y_\omega = \mathcal{B}_{y_{d,q}}(\mathcal{M}_{\phi}(x)) \text{.}
\end{equation}
At each optimization step, the Meta Learner optimizes $\mathcal{M}_{\phi}(x)$ using a specific slice $y_\omega$, encouraging the construction of a discriminative auxiliary embedding among different label combinations.
This mechanism ties $f_{\theta}^\omega$ to the joint semantic information of image quality and disease diagnosis. 
By working in conjunction with meta-auxiliary learning, it contributes to adaptively and effectively capturing the interplay between image quality and disease diagnosis, ultimately leading to improved model performance.

\textbf{Meta-auxiliary learning.} 
Meta Learner is designed to optimize Task Net to construct $f_{\theta}^\omega$ effectively for diagnostic purposes.
This is essentially a ``learning-to-learn'' problem, for which we adopt meta-auxiliary learning to optimize the effectiveness of $f_{\theta}^\omega$ by providing a beneficial and desired $y_\omega$ \cite{liu2019self}. 
The idea is that training Task Net on $y_\omega$ should result in improved performance.
Therefore, Meta Learner $\mathcal{M}_{\phi}$ should optimize $\phi$ in the direction that minimizes the objective function $\mathcal{L}_{\theta}$ of Task Net $\mathcal{M}_{\theta}$, i.e., $\arg \min_{\phi} \mathcal{L}_{\theta}$. 
To accomplish this, we first perform a one-step pseudo update with learning rate $\alpha$ to simulate the effect of $y_\omega$ on $\theta$ as
\vspace{-4mm}
\begin{equation}
    \begin{split}
    \tilde{\theta} := \theta - \alpha \nabla_{\theta} \big[ \mathcal{L}_D \big( \mathcal{M}_{\theta}^d (x), y_d\big)+ \mathcal{L}_Q \big( \mathcal{M}_{\theta}^q (x), y_q\big) \\
     + \mathcal{L}_\Omega \big( \mathcal{M}_{\theta}^\omega (x), y_\omega \big) \big] \text{.}
    \end{split}
\end{equation}
Then, we use the second derivative trick to update $\phi$, similar to \cite{finn2017model,zhang2018fine}:
\vspace{-1mm}
\begin{equation}
    \begin{split}
    \phi := \phi - \beta \nabla_{\phi} \big[ \mathcal{L}_D \big( \mathcal{M}_{\tilde{\theta}}^d (x), y_d\big) + \mathcal{L}_Q \big( \mathcal{M}_{\tilde{\theta}}^q (x), y_q\big) \\
     + \mathcal{R}\big(\mathcal{B}_{y_{d,q}}(\mathcal{M}_{\phi}(x)) \big) \big] \text{,}
    \end{split}
\end{equation}
where $\beta$ is the learning rate of Meta Learner, and $\mathcal{R}(\cdot)$ is a regularization term to avoid auxiliary label embedding collapse by increasing the entropy of $\mathcal{M}_{\phi}(x)$  \cite{liu2019self, liu2022auto}.

Meta Learner leverages gradient information to adaptively generate $\mathcal{M}_{\phi}(x)$ to assist Task Net in constructing $f_{\theta}^\omega$.
Minimizing $\mathcal{L}_\theta$ allows Meta Learner to provide useful $\mathcal{M}_{\phi}(x)$ for solving diagnosis and quality assessment tasks \cite{zhang2018fine}, which in turn ensures that $f_{\theta}^\omega$ contains the desired semantic information derived from learning $y_\omega$.
Furthermore, this meta-optimization procedure takes into account the MAB where $f_{\theta}^\omega$ is utilized for diagnosis, which further enhances the effectiveness of $f_{\theta}^\omega$. 

\begin{table*}[!tbp]
    \renewcommand\arraystretch{1.05}
    \centering
        \caption{\small{Statistics for lung disease diagnosis datasets (left group) and ophthalmic disease diagnosis datasets (right group).}}
        \label{tab:DR_statis}
        \resizebox{1.0\textwidth}{!}{%
        \scalebox{0.69}{
        \begin{tabular}{c|ccc|cccc||c|ccc|ccc|cccc}
            \hline
            \hline
              \multicolumn{1}{c|}{Dataset} &\multicolumn{3}{c|}{CT-IQAD} &\multicolumn{4}{c||}{CXR-IQAD} &\multicolumn{1}{c|}{} &\multicolumn{3}{c|}{DRAC} &\multicolumn{3}{c|}{DeepDR} &\multicolumn{4}{c}{EyeQ} \\
            \hline

            Label &ALL & HQ & LQ &ALL & HQ & LQ-C & LQ-A & Label &ALL & HQ & LQ &ALL & HQ & LQ &ALL & HQ & LQ-U & LQ-P  \\
            \hline
            \hline
            Normal &1,149 &800 &349 &2,582 &792  &791 &999 & NoDR &545 &444 &101 &914 &532 &382 &20,680 &12,308 &4,553 &3,747  \\
            Covid-19 &1,197 &800  &397  &- &- &- &- & NPDR &344 &303 &41  &974 &506  &468 &7,533 &4,405  &1,690 &1,438  \\
            Pneumonia &- &- &- &5,704 &2,137 &2,136 &1,431 & PDR &108 &93  &15 &112 &50  &62  &558 &104  &191  &353  \\
            \hline
            ALL &2,346 &1,600 &746  &8,286 &2,929 &2,927 &2,430 & ALL &997 &840  &157 &2,000  &1,088  &912 &28,789 &16,817  &6,434  &5,538  \\
            \hline
            \hline
            \end{tabular}
    }}
    \vspace{-2mm}
\label{tab:DR_Dataset}
\end{table*}

\begin{table*}[!tbp]
    \renewcommand\arraystretch{1.1}
    \centering
        \caption{\small{Comparison with state-of-the-art approaches in disease diagnosis.}}
        \label{tab:comparisonsfundus}
        \resizebox{1.0\textwidth}{!}{%
        \scalebox{0.69}{
        \begin{tabular}{c|ccc|ccc|ccc|ccc|ccc|ccc}
            \hline
            \hline
             Dataset &\multicolumn{3}{c|}{CT-IQAD} &\multicolumn{3}{c|}{CXR-IQAD} &\multicolumn{3}{c|}{DRAC} &\multicolumn{3}{c|}{DeepDR} &\multicolumn{3}{c|}{EyeQ} &\multicolumn{3}{c}{Avg.} \\
             \hline
            Metrics& AUC &ACC & F1 & AUC &ACC & F1&  AUC &ACC & F1&  AUC &ACC & F1 & AUC &ACC & F1 & AUC &ACC & F1  \\
            \hline
            \hline
            MMCNN \cite{zhou2018multi} &86.57 &86.54 &86.51 &89.13 &89.07 &91.82 &82.41 &77.20 &66.18 &76.39 &79.50 &54.40 &69.36 &73.22 &35.48 &80.77 &81.11 &66.88 \\
            BIRA-Net \cite{zhao2019bira} &88.46 &88.46 &88.21 &89.88 &90.76 &93.23 &84.00 &77.46 &67.00 &80.67 &78.50 &64.10 &80.13 &72.50 &56.42 &84.63 &81.54 &73.79 \\
            GREEN \cite{liu2020green} &92.54 &92.52 &92.44 &90.06 &91.30 &93.67 &84.37 &73.83 &63.03 &81.14 &76.50 &\textbf{64.88} &81.23 &77.21 &59.08  &85.87 &82.27 &74.64 \\
            CABNet \cite{he2020cabnet} &91.01 &91.03 &90.79 &88.38 &89.73 &92.51 &81.94 &72.28 &59.07 &82.39 &77.75 &55.97 &79.21 &76.95 &53.48  &84.59 &81.55 &70.36 \\
            \hline
            Mixstyle \cite{zhou2021domain} &91.06 &91.03 &90.99 &88.85 &90.52 &93.14 &85.03 &73.32 &60.91 &81.89 &73.50 &60.34 &82.10 &74.92 &\underline{61.25}  &85.79 &80.66 &73.33 \\
            AugMix \cite{hendrycks2019augmix} &89.17 &89.10 &89.26 &88.21 &89.79 &92.58 &84.94 &70.47 &57.66 &\underline{84.79} &\underline{80.00} &54.74 &77.38 &77.00 &44.26  &84.90 &81.27 &67.70 \\
            DDAIG \cite{zhou2020deep} &92.72 &92.74 &92.54 &88.19 &90.28 &93.00 &84.95 &77.72 &62.91 &79.40 &75.75 &51.66 &70.71 &74.64 &37.84 &83.19 &82.23 &67.59\\
            Mixup \cite{zhang2017mixup} &92.12 &92.10 &92.04 &88.98 &90.70 &93.28 &82.41 &72.80 &62.84 &81.69 &76.00 &63.57 &82.35 &\underline{77.74} &\textbf{63.07} &85.51 &81.87 &\underline{74.89} \\
            \hline
            QGNet\cite{zhou2018fundus} &92.71 &92.74 &92.50 &89.51 &90.70 &93.22 &84.38 &63.73 &47.80 &84.21 &73.50 &\underline{64.66} &82.40 &74.73 &60.36  &86.64 &79.08 &71.71 \\
            MAXL\cite{liu2019self} &93.35 &93.38 &93.16 &87.99 &89.86 &92.66 &\underline{87.24} &\underline{77.72} &\underline{69.20} &81.59 &71.25 &57.58 &80.64 &73.10 &58.81  &86.16 &81.06 &74.28\\
            CANet\cite{li2019canet} &93.36 &93.38 &93.19 &\underline{90.51} &91.49 &93.78 &86.70 &75.13 &64.38 &81.83 &77.50 &62.74 &81.89 &77.20 &60.26  &\underline{86.86} &\underline{82.94} &74.87 \\
            MT-Net\cite{chen2019multi} &92.56 &92.52 &92.51 &90.07 &90.58 &93.05 &84.63 &73.06 &61.89 &82.40 &76.75 &64.51 &\underline{82.66} &75.97 &59.92  &86.46 &81.78 &74.38 \\
            MTMR-Net \cite{liu2019multi} &\underline{94.46} &\underline{94.44} &\underline{94.37} &89.69 &\underline{91.61} &\underline{93.96} &85.79 &76.42 &65.12 &82.30 &63.75 &53.61 &79.71 &76.57 &60.50  &86.39 &80.56 &73.51 \\
            DETACH \cite{che2022learning} &92.53 &92.52 &92.41 &89.52 &90.64 &93.16 &84.81 &76.94 &65.24 &83.61 &71.75 &62.14 &81.68 &74.80 &59.64  &86.43 &81.33 &74.52 \\
            \hline
            \textbf{MKCNet (Ours)} &\textbf{95.73} &\textbf{95.73} &\textbf{95.65} &\textbf{90.71} &\textbf{91.91} &\textbf{94.12} &\textbf{88.68} &\textbf{82.38} &\textbf{73.87} &\textbf{86.58} &\textbf{80.00} &54.73 &\textbf{83.83} &\textbf{78.01} &59.68  &\textbf{89.11} &\textbf{85.61} &\textbf{75.61} \\
            \hline
            \hline
            \end{tabular}
    }}
    \vspace{-3mm}
\label{tab:comparion_eye}
\end{table*}

\section{Experiment}
\subsection{Dataset and Implementation Details}
\textbf{Dataset.} We employ five widely-used public datasets from four imaging modalities for three diagnosis tasks.
In the data preprocessing step, all images were resized to $256\times256$ and then individually normalized to zero mean and unit variance in intensity values.
Table \ref{tab:DR_Dataset} presents the statistics of the datasets, and their settings are briefly described as follows:
\textbf{DRAC}\cite{qian2023drac}:
OCTA images are graded into three levels, standing for no DR, non-proliferative DR (NPDR), and proliferative DR (PDR), respectively. 
Each image is annotated with HQ or LQ.
\textbf{DeepDR}\cite{liu2022deepdrid}:
Same with DRAC, fundus images are graded into three levels of DR and two levels of image quality.
\textbf{EyeQ}\cite{fu2019evaluation}:
The diagnosis annotation of fundus images is the same as DeepDR, while the quality of images is labeled as good (HQ), usable (LQ-U), and poor (LQ-P).
\textbf{CT-IQAD}:
We combine computed tomography images from COVID-X \cite{zhao2020covid} and SARS \cite{soares2020sars} to make up this dataset.
Since COVID-X images are collected in the wild, e.g., downloaded from papers, their quality cannot be guaranteed, and thus we label these images LQ and others are HQ.
\textbf{CXR-IQAD}:
This dataset is sampled from child chest X-ray images of CXR-P \cite{kermany2018identifying} and adult CXR images of CXR14 \cite{wang2017chestx}.
The diagnosis label is Normal and Pneumonia.
We downsample images with a bicubic kernel to simulate low-dose CXR images \cite{xu2020low}.
Thus, images have three parts, i.e., child images (HQ), low-dose child images (LQ-C), and low-dose adult images (LQ-A).
During training, we use the original data splits for DeepDR and DRAC. 
We split EyeQ into three parts: 60\% for training, 10\% for validation, and 30\% for testing.
As for the rest of the datasets, we divide them into 70\% for training, 10\% for validation, and 20\% for testing. 
To evaluate the performance fairly, we used three metrics: the area under the ROC curve (AUC), accuracy (ACC), and macro F1-score (F1).


\textbf{Implementation.}
We adopt VGG16 \cite{simonyan2014very} as backbones in the experiment, and the focal loss \cite{lin2017focal} and entropy loss for training.
The learning rate and weight decay are kept the same for all datasets at 0.01 and 0.0005, respectively, and we adopt SGD as the optimizer.
The number of training epochs is set as 100 for EyeQ and 200 for the rest of the others.
We followed the train-validation-test paradigm to report results for all datasets except DRAC, i.e., we selected the best models in validation and evaluated them on the test set.
For DRAC, we reported the last epoch result because its official data split does not include a validation set.
Further details about this sub-section can be found in the appendix.

\subsection{Comparison with Other Methods}

\textbf{Experiment setting.} 
We compare MKCNet with recent state-of-the-art methods from three groups: ophthalmic disease diagnosis (ODS) methods \cite{zhou2018multi, zhao2019bira, liu2020green, he2020cabnet}, multi-task and auxiliary learning (MTAL) methods \cite{zhou2018fundus,li2019canet,chen2019multi,liu2019multi,liu2019self,che2022learning}, and other adoptable methods \cite{zhang2017mixup,zhou2021domain,zhou2020deep,hendrycks2019augmix}.
The chosen methods can be adopted in IQAD without or with only simple modifications, and their brief descriptions are in the appendix. 

\textbf{Comparison of results.} 
Columns 3-5 of Table \ref{tab:comparion_eye} show the quantitative results of ophthalmic disease diagnosis.
MKCNet outperforms other methods in overall performance and improves at least two of the AUC, ACC, and F1 metrics on all datasets.
Typically, ODS methods do not consider using image quality information and only design modules for achieving good performance on HQ images. 
Thus, most of them perform non-ideally on the IQAD problem, where numerous LQ images exist.
For example, GREEN \cite{liu2020green} considers relationships among categories to achieve a robust diagnosis, yet it suffers in IQAD. 
It is because image degradations significantly disturb the model to construct the foundations of relationship exploration among categories, i.e., correlations between images and labels.
We modified methods from the MTAL set to learn image quality information as an auxiliary task. 
As observed in the last column of Table \ref{tab:comparion_eye}, a rough sense is that the average performance of MTAL methods outperforms OSD methods.
Such advance illustrates the importance of image quality information in the IQAD problem.
Nevertheless, MKCNet outperforms these MTAL methods due to its explicit image quality-assist modules and meta-learning techniques that bypass the problem of coarse quality labels and provide sufficient information to diagnose problems. 
It is worth mentioning that CANet explicitly explores the internal relationships between two diseases, and we modified it to explore the assistance of quality assessment in diagnosis. 
Although the coarse quality label only contributes scant information, CANet still surpasses other methods, demonstrating the importance of explicit utilization mechanisms. 
We also tested MKCNet on two different lung disease diagnosis tasks on two different imaging modalities to verify its generalization ability. 
Columns 1-2 of Table \ref{tab:comparion_eye} present the quantitative results of lung disease diagnosis. 
As expected, MKCNet significantly outperforms other methods even when facing diagnosis tasks on another organ with different imaging modalities, owing to its unique modeling for the IQAD problem. 
MKCNet outperforms the methods ranked second with a significant margin, achieving an average AUC, ACC, and F1 improvement of \textbf{2.25}\%, \textbf{2.67}\%, and \textbf{0.72}\%, respectively, highlighting its effectiveness.

\subsection{Ablation Study}

\begin{figure}[t]
    \centering
    \includegraphics[width=\linewidth]{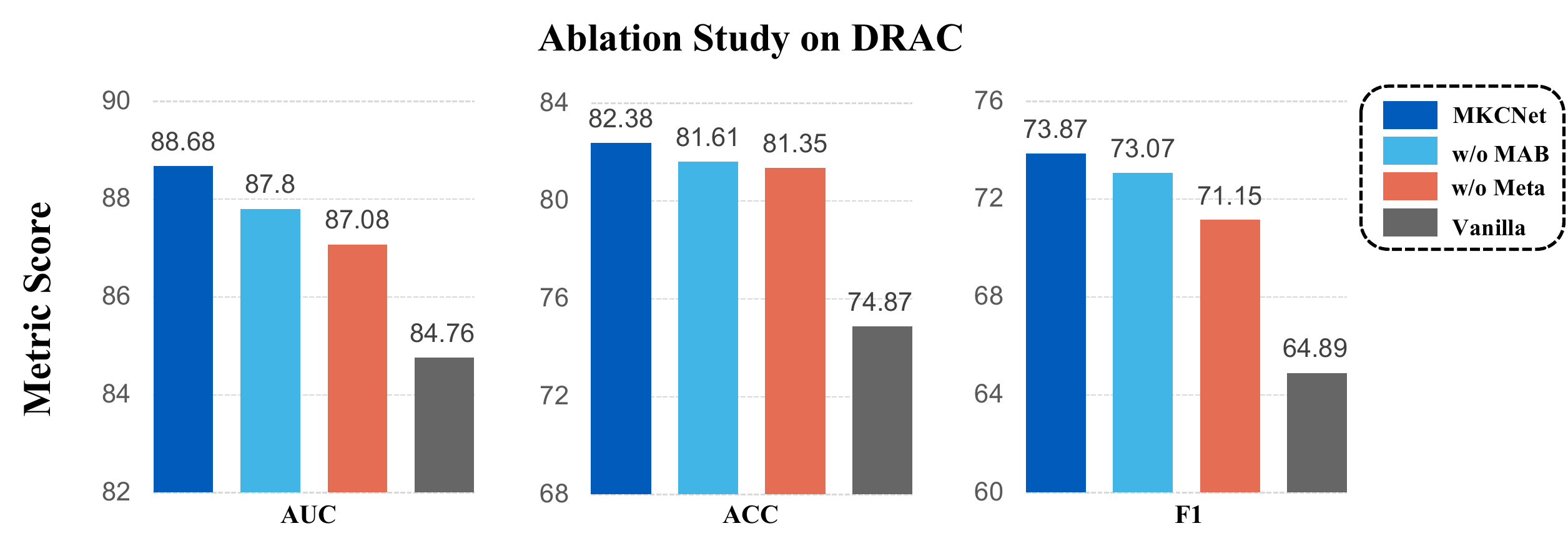}
    \caption{\small{Ablation study for the effect of proposed components.}}
    \label{fig:abaltion}
\vspace{-2mm}
\end{figure}

\textbf{Contribution of each component.}
We first validate the effects of the two critical components in our method by removing Meta Learner (denoted as w/o Meta, which will leverage $f_\theta^q$ to assist diagnosis) and removing the explicit quality information utilization mechanism (denoted as w/o MAB, where Meta Learner will still optimize Task Net).
We also include a baseline model with the same basic experimental setting as ours, denoted as Vanilla.
As shown in Figure \ref{fig:abaltion}, removing either component results in decreased performance of MKCNet on all three metrics, but the performance is still better than Vanilla. 
This result is reasonable and suggests that these two components play complementary roles in addressing the IQAD problem.

\begin{figure}[t]
    \centering
    \includegraphics[width=\linewidth]{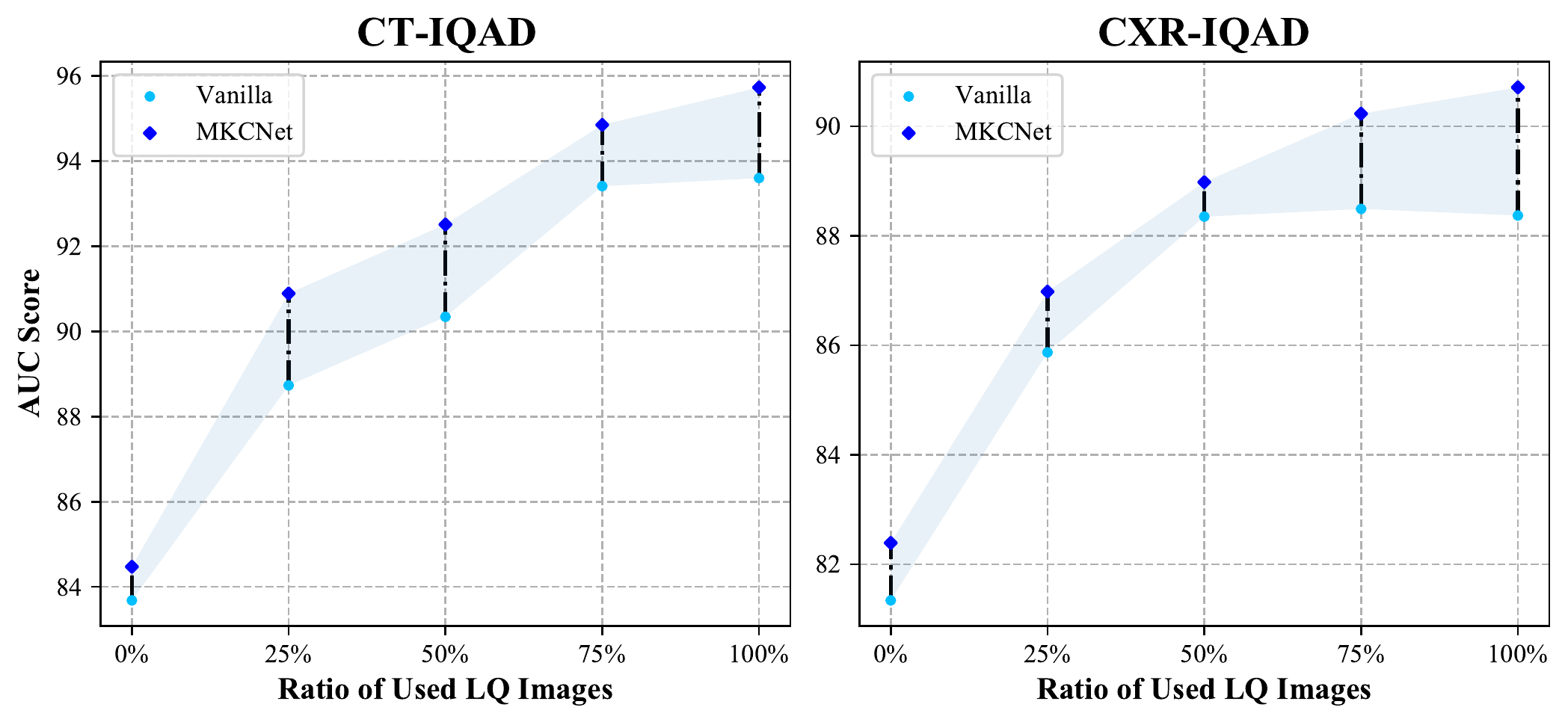}
    \caption{\small{Performance of MKCNet and Vanilla on different ratios of LQ images used in training on CT-IQAD and CXR-IQAD.}}
    \label{fig:IQ}
\vspace{-4mm}
\end{figure}

\begin{figure}[t]
    \centering
    \includegraphics[width=\linewidth]{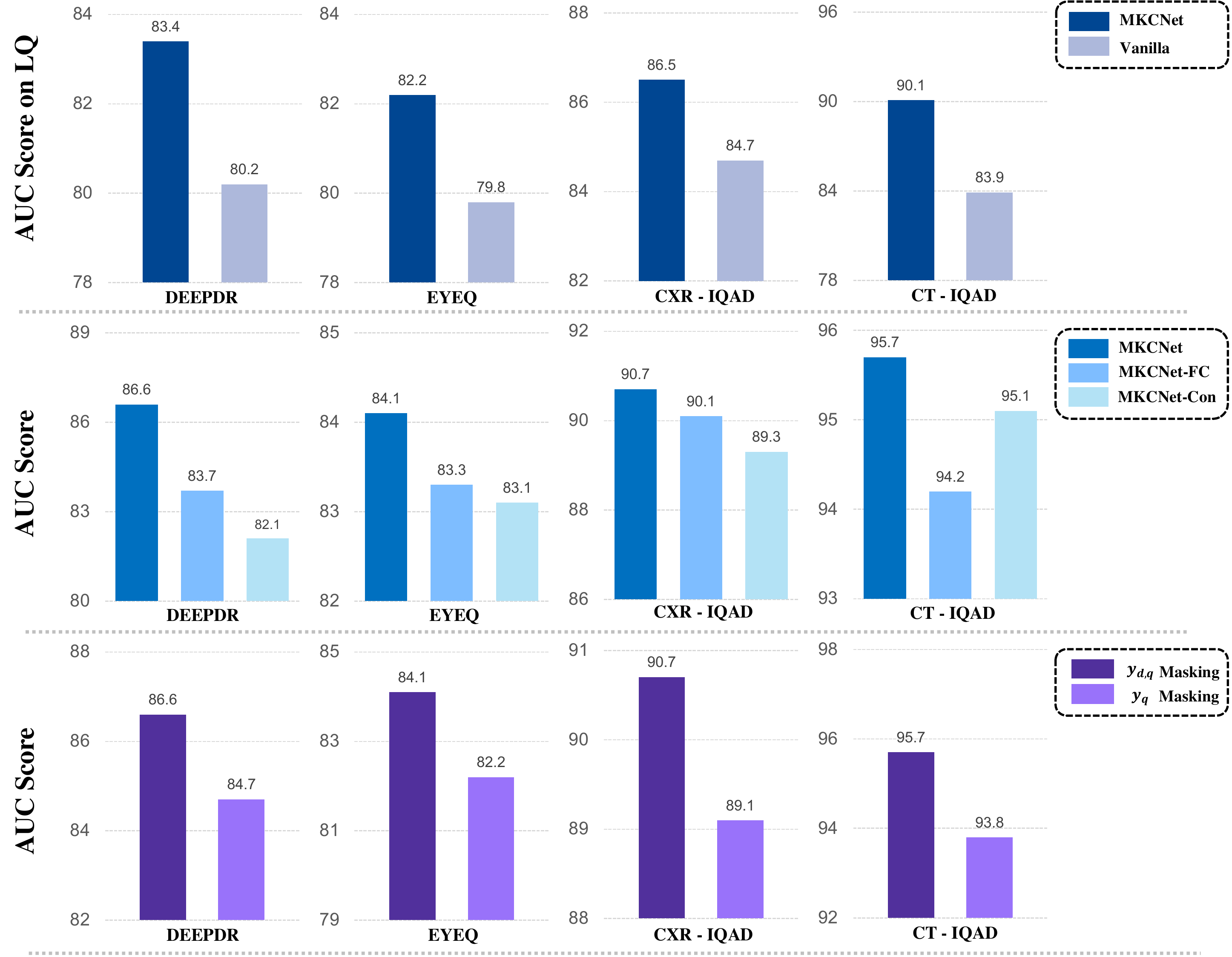}
   \caption{\small{Histogram of AUC score analysis on four datasets.
   \textbf{Top}: Comparison on LQ images.
   \textbf{Middle}: The influence of Task Net design.
   \textbf{Bottom}: The impact of different quality information.}}
    \label{fig:cotask}
\vspace{-4mm}
\end{figure}

\textbf{Influence of LQ images.}
We evaluated the effectiveness of MKCNet in handling different numbers of LQ images during training to demonstrate its generalization ability, as illustrated in Figure \ref{fig:IQ}.
To conduct this experiment, we divided LQ images into four equal parts and trained Vanilla and MKCNet with 0\%, 25\%, 50\%, 75\%, and 100\% of LQ images while retaining the same number of HQ images.
As expected, the performance of both Vanilla and MKCNet increased with the addition of LQ images, but the performance gap between the two methods became more pronounced.
Furthermore, we compared the performance of MKCNet with Vanilla on LQ images.
As shown at the top of Figure \ref{fig:cotask}, MKCNet achieved a significant performance improvement over Vanilla.
These experiments highlight the importance of learning from LQ images and demonstrate that MKCNet can more effectively leverage LQ images by utilizing quality labels compared to Vanilla.

\textbf{Effect of designs in Task Net.}
We conducted an in-depth analysis to demonstrate the effectiveness of designs in Task Net. 
First, we discussed modifications to investigate the need for the GAB and the way of explicit utilization mechanism.
We replaced the GAB with fully connected layers and kept other components unchanged, denoted as MKCNet-FC. 
Additionally, we designed MKCNet-Con, which diagnosed based on the concatenation of $f_\theta^d$ and the entire meta co-embedding feature $f_\theta^\omega$, without filtering diagnosis-irrelevant channels. 
As shown in the middle of Figure \ref{fig:cotask}, the performance drop of these variants indicates the effectiveness and necessity of these designs. 
The results suggest two critical issues: a) the GAB contributes to desired feature representation construction, and b) designing a practical utilization mechanism is crucial for IQAD.

\begin{figure}[t]
    \centering
    \includegraphics[width=\linewidth]{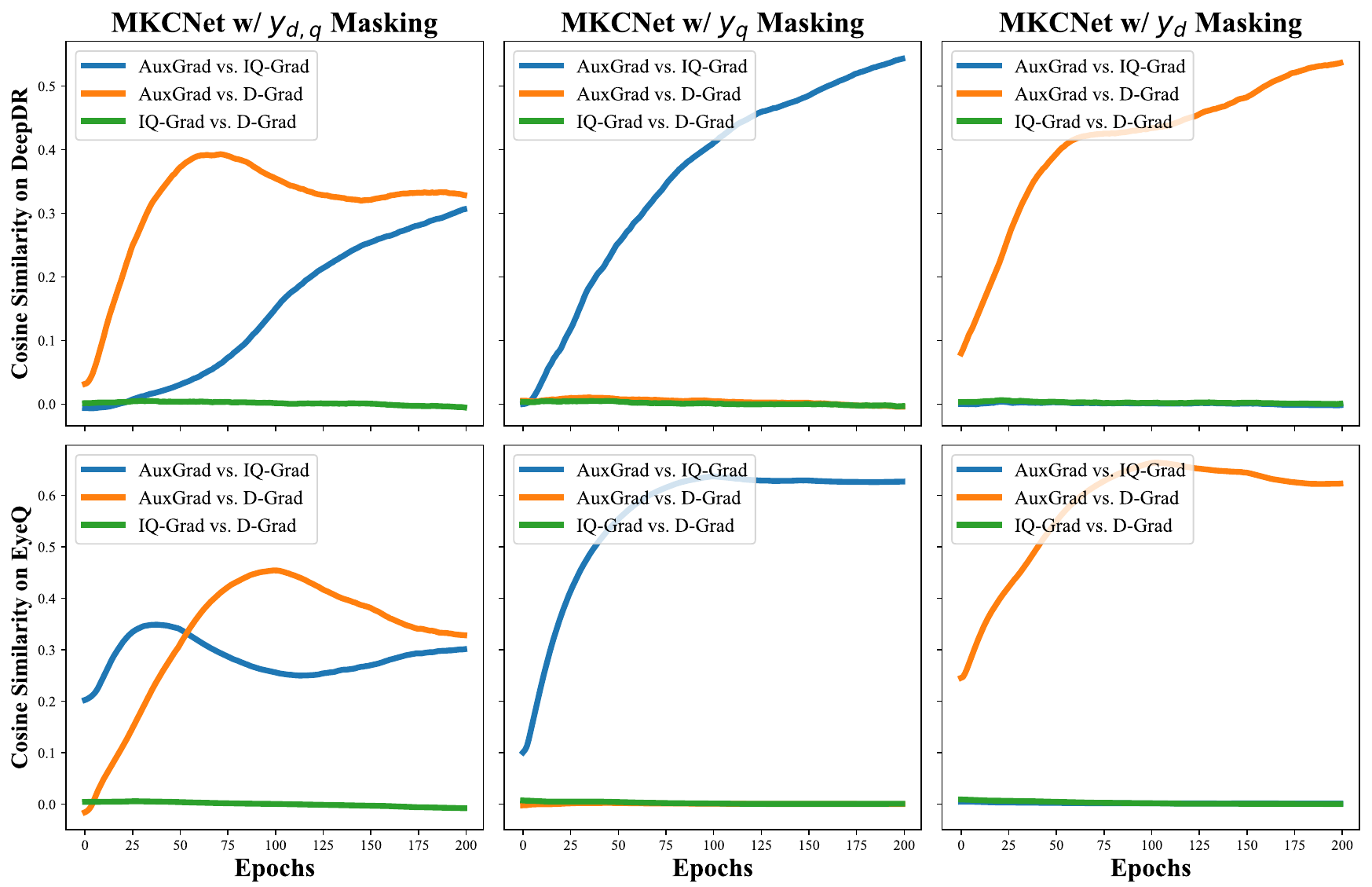}
   \caption{\small{Gradient cosine similarity analysis on Task Net.}}
    \label{fig:cos}
\vspace{-3mm}
\end{figure}

\textbf{Analysis of Meta Learner.}
We first conducted experiments to verify the effectiveness of joint-encoding masking, which is used to select the auxiliary label embedding $y_w$. 
We compared the results of using different masking strategies and found that only using the image quality label for masking decreased performance significantly, as shown at the bottom of Figure \ref{fig:cotask}. 
This illustrates the importance of joint-encoding masking, which uses joint label encoding to ensure that the semantics of $y_\omega$ correlate with both image quality and disease diagnosis.
We further analyzed the cosine similarity among the auxiliary label embedding loss gradient (AuxGrad), diagnosis loss gradient (D-Grad), and quality assessment loss gradient (IQ-Grad) on the shared backbone in Task Net. 
Figure \ref{fig:cos} shows that $y_\omega$ provides information related to both diagnosis and image quality, while there is little similarity among the loss gradients of the disease diagnosis and image quality assessment branch. 
We also observed that the label type used in the embedding masking affects the auxiliary loss gradient and its correlation with the corresponding task. 
If we use the diagnosis label to obtain $y_\omega$, the gradient of that learning branch will be highly related to disease diagnosis but uncorrelated with image quality assessment, and vice versa.
These analysis support the effectiveness of knowledge co-embedding features, which contain the desired knowledge correlated with both quality assessment and disease diagnosis.

\begin{figure}[t]
    \centering
    \includegraphics[width=\linewidth]{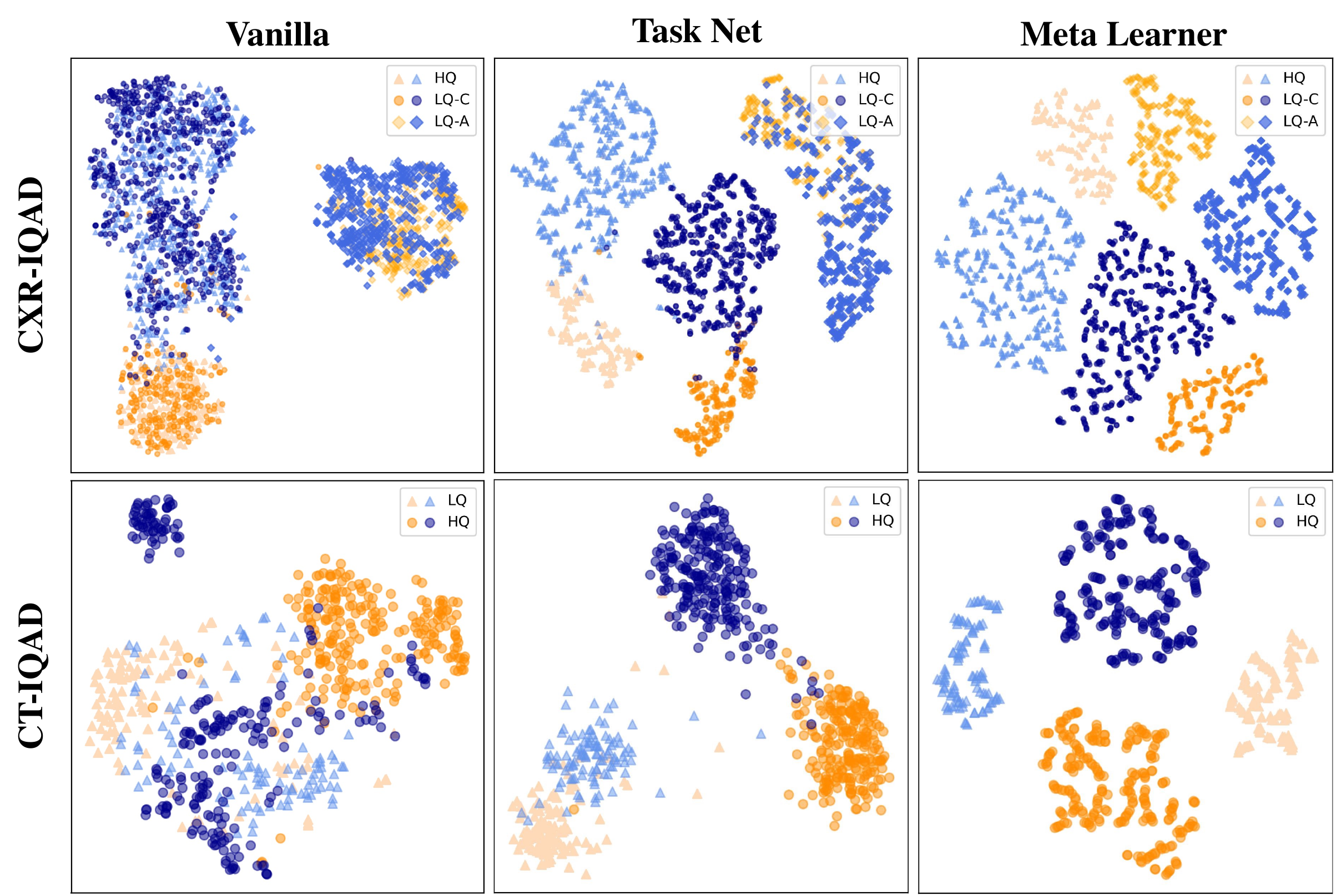}
   \caption{\small{The t-SNE analysis. Similar colors and distinct shapes indicate the same diagnosis and different quality labels. }}
    \label{fig:tsne}
\vspace{-4mm}
\end{figure}

\textbf{Importance of auxiliary label embedding $y_\omega$.}
We conducted t-SNE analysis on the output vector $\mathcal{M}_\phi(x)$ of Meta Learner and the final diagnosis feature $f_{\theta}^{d^*}$ to visually understand the effect of auxiliary label embedding.
As shown in Figure \ref{fig:tsne}, Task Net successfully learned discriminative features for disease diagnosis, and these features tended to be clustered based on image quality and disease diagnosis categories.
However, Vanilla only focused on diagnosis labels and obscured features of images with different image qualities.
This indicates the effectiveness of MKCNet in leveraging diagnosis and quality labels, i.e., it comprehensively considers both diagnosis and quality information to diagnose.
From the visualization of $\mathcal{M}_\phi(x)$, we can observe two facts: a) these output vectors are clustered by the joint encoding of image quality and diagnosis, and b) feature points are dispersed inside each cluster.
The first fact illustrates the justification of Meta Learner, which learns fused and discriminative information regarding image quality and disease diagnosis.
The latter fact illustrates that the auxiliary label embedding is diverse with images and may provide adaptively helpful information to $f_\theta^\omega$.

\begin{figure}[t]
    \centering
    \includegraphics[width=\linewidth]{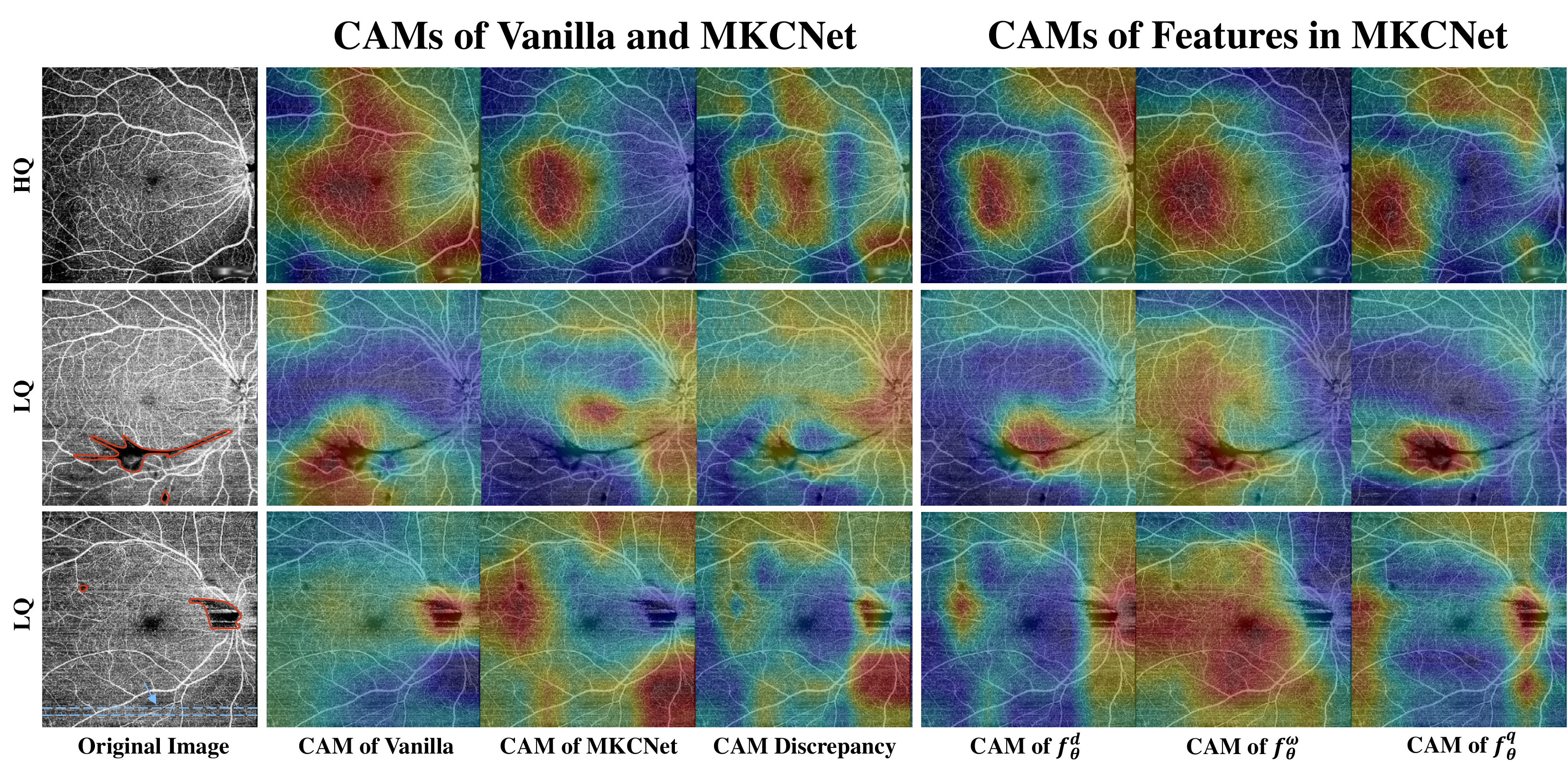}
    \caption{\small{Qualitative analysis via CAM visualization.}}
    \label{fig:cam}
\vspace{-4mm}
\end{figure}

\textbf{Qualitative analysis.} 
Finally, we performed class activation mappings (CAMs) using \cite{zhou2016learning} to qualitatively investigate how MKCNet performs on DARC with varying image qualities. 
The results, shown in Figure \ref{fig:cam}, demonstrate that both Vanilla and MKCNet effectively attend to the HQ images. 
However, Vanilla fails to deal with degradations in the images, resulting in misleading signs. 
In the second row of the figure, for instance, Vanilla focuses on the fake vessel disappearing in the low-signal area caused by threshold degradation. 
In the third row, it focuses on the false abnormality of the optic disc due to segment degradation and low-signal area. 
In contrast, MKCNet is more robust in dealing with such degradations and ignores misleading clues while capturing the foveal avascular zone or vessel density as desired. 
Moreover, the CAM visualization of features in MKCNet demonstrates how quality information can assist in diagnosis. 
In the second and third rows, $f_\theta^d$ initially perceives the area of interest as similar to the area of interest of $f_\theta^q$; i.e., focuses on degradations, which indicates that it has been misled. 
However, after leveraging information from $f_\theta^\omega$, it refocuses on the vessel area density or foveal avascular zone. 
This illustrates the importance of quality information and the effectiveness of our model in providing assistance or calibration to the diagnosis.

\section{Conclusion}
In this paper, we address the significant yet often overlooked problem of IQAD by reconsidering the value of LQ images and quality labels. 
To achieve this, we propose a novel MKCNet consisting of two subnets.
Task Net explicitly explores the potential benefits of quality labels in diagnosis by leveraging knowledge co-embedding features, while Meta Learner learns to optimize these features to ensure their desired effectiveness and semantics. 
The experiments show the superior performance of our method on five datasets spanning four widely used medical imaging modalities. 
Our method effectively overcomes the challenges of leveraging quality labels, offering a more practical and cost-effective solution than conventional approaches. 
By assessing the value of LQ images and quality labels, we highlight their importance in diagnosis tasks. 
Furthermore, it is expected to pave the way for future research in leveraging quality labels and LQ images, stimulating the development of robust, accurate, and practical diagnostic models that can effectively handle real-world challenges in clinical settings.

\vspace{5pt}
\textbf{Acknowledgement.} 
This work was supported by Shenzhen Science and Technology Innovation Committee Fund (Project No. SGDX20210823103201011 and JCYJ20180507182410327), Hong Kong Innovation and Technology Fund (Project No. ITS/028/21FP) and HKUST-WeBank Joint Lab (Project No. WEB19EG01-Q).
We would like to express our sincere gratitude to Huajun Zhou, Haibo Jin, Hao Jiang, Zhengrui Guo, Dawei Yang, Jingran Su, Kang Zhou, Lai Tian, and  for their valuable discussions.

\clearpage
{\small
\bibliographystyle{unsrt}
\bibliography{mybib}
}



\end{document}